\documentclass[11pt]{article}

\usepackage{amssymb}
\usepackage{color}
\usepackage{epsfig,amssymb,amsfonts,amsmath,graphicx,dsfont,cite,xfrac}
\usepackage{authblk}
\usepackage{subcaption}
\definecolor{mygray}{gray}{0.5}

\usepackage{cite}
\usepackage[colorlinks=true,linkcolor=blue,citecolor=red]{hyperref}

\newcommand{\be}{\begin{equation}}
\newcommand{\ee}{\end{equation}}
\newcommand{\bea}{\begin{eqnarray}}
\newcommand{\eea}{\end{eqnarray}}

\usepackage{xfrac,amssymb,amsfonts,amsmath}
\usepackage{tikz,pgfplots}
\usetikzlibrary{positioning}
\usetikzlibrary{matrix}

\title{Exact solutions for time-dependent non-Hermitian oscillators: classical and quantum pictures}

\author[${1}$]{Kevin Zelaya}
\author[${2}$]{Oscar Rosas-Ortiz}

\affil[${1}$]{\footnotesize Nuclear Physics Institute, Czech Academy of Science, 250 68 \v{R}e\v{z}, Czech Republic}
\affil[${2}$]{\footnotesize Physics Department, Cinvestav, AP 14-740, 07000 M\'exico City, Mexico}

\date{}
\begin{document}

\maketitle

\begin{abstract}
We associate the stationary harmonic oscillator with time-dependent systems exhibiting non-Hermiticity by means of point transformations. The new systems are exactly solvable, with all-real spectrum, and transit to the Hermitian configuration for the appropriate values of the involved parameters. We provide a concrete generalization of the Swanson oscillator that includes the Caldirola-Kanai model as a particular case. Explicit solutions are given in both, the classical and quantum pictures.
\end{abstract}

\section{Introduction}

During the last decades, there has been an increasing interest in non-Hermitian structures and their implications in quantum theory. Nevertheless, the subject was overlooked for long time because the reality of the spectrum is not granted a priori for non-Hermitian models. A surprising breakthrough was offered by the demonstration that parity-time (PT) symmetry implies real spectrum~\cite{Ben98}, which stimulated the systematic search of PT-symmetric systems in quantum mechanics \cite{Lev20} (see also papers in the special issues \cite{Gey06,Fri08}). Further improvements shown that $PT$-symmetry is not a necessary condition for the spectrum reality~\cite{Mos02,Mos10}, a fact confirmed in diverse models of non-Hermiticity \cite{Ros15,Ros18,Bla19,Zel20a,Zel20d}, where the imaginary part of a wide class of complex-valued potentials lead to balanced gain and loss probability without the necessity of PT-symmetry. The latter property opens new possibilities in optical design, where the gain-loss manipulation includes but is not limited to PT-Symmetry \cite{Gbu18}. Important theoretical achievements within the PT-symmetric formulation \cite{Gan07} envisioned the experimental observation of the phenomenon in optics \cite{Rut10}.

Another interesting problem deals with nonstationary systems in quantum mechanics, which finds exciting applications in plasma physics~\cite{Mih19} as well as in the design of electromagnetic traps for charged particles~\cite{Pau90,Com86,Pri83,Gla92,Mih09,Mih21}. As time-dependent systems obey dynamical equations that cannot be reduced to eigenvalue problems in general, finding solutions could mean a formidable consumption of computing resources. Two simplifications are notable, systems obeying adiabatic evolutions~\cite{Sch02} and the Lewis-Riesenfeld (parametric) oscillator~\cite{Lew69}. In the former case the time-evolution of the system is slow enough, so the Hamiltonian tolerates time-dependent eigenvalues for instantaneous eigenvalue equations~\cite{Boh03}. In turn, the parametric oscillator is found to have a constant of motion that defines the appropriate eigenvalue equation \cite{Lew69}. The latter result motivated the systematic research of quantum invariants \cite{Dod95,Gue15b,Zel20c,Dod21}, with applications in the construction of time-dependent wave-packets \cite{Cas13,Cru15,Cru16,Zel20b,Cru17,Una18,Zel19,Cru20,Zel21a}, Darboux transformations \cite{Bag95,Zel17,Cen19,Raz19}, and two-dimensional photonic systems \cite{Con19}, among other.

The Swanson oscillator~\cite{Swa04} is a very peculiar system that combines both profiles since it is non-Hermitian and time-dependent. Formulated to study transitions of probability amplitudes that are generated by non-unitary time evolutions, the model developed by Swanson is revisited and studied in different branches of physics and mathematical physics \cite{Mid11,Gre15,Bag15a,Bag15b,Dou21,Fri21,Lim21}. Quite remarkably, the Swanson Hamiltonian can be connected with the Hamiltonian of the harmonic oscillator by the appropriate rotation in configuration space \cite{Bag21}, which clarifies the solvability of the model.

In this work we use point transformations \cite{Ste93} to associate the stationary harmonic oscillator with a time-dependent system that exhibits non-Hermiticity and all-real spectrum. We have already applied the method to study parametric oscillators in the Hermitian regime \cite{Zel20c}, where we also shown that the construction of coherent states is feasible for such systems. Here, we show that the appropriate point transformation yields a wide family of time-dependent systems that may be chosen to be Hermitian or non-Hermitian, according with the involved parameters. Associating these systems with the stationary harmonic oscillator we obtain exactly solvable models that recover a diversity of oscillators in both, Hermitian and non-Hermitian configurations.

The organization of the paper is as follows. In Sec.~\ref{sec:hosw} we introduce the main concepts and provide the space of solutions for the non-Hermitian, time-dependent systems in terms of the well known solutions of the harmonic oscillator. Concrete expressions are given for the Hamiltonian of such systems. In Sec.~\ref{applica} we show the applicability of our model by solving the dynamical law of a generalization of the Caldirola-Kanai oscillator in both, classical and quantum pictures. We have added App.~\ref{ApA}, where the point transformation theory is summarized.
 
\section{Generalized Oscillators}
\label{sec:hosw}

Let us assume that the operator
\begin{equation}
\tfrac{1}{\hbar w_{0}} H_{\operatorname{sw}}=\alpha_{2}(t) \hat{a}^{\dagger 2} +\beta_{2}(t)\hat{a}^{2}+\theta(t) \{ \hat{a},\hat{a}^{\dagger} \}+\alpha_{1}(t)\hat{a}^{\dagger}+\beta_{1}(t)\hat{a}\, ,
\label{sw1}
\end{equation}
rules the dynamical law of a time-dependent quantum system. The constant $w_0 >0$ is written in units of frequency. Hereafter $\hat a$ and $\hat a^{\dagger}$ represent the boson ladder operators fulfilling $[\hat a, \hat a^{\dagger} ]= \mathbb I$, with $\mathbb I$ the identity operator in the Hilbert space spanned by the Fock basis $\{ \vert n \rangle, n=0,1,2, \ldots\}$. The symbols $[ \cdot, \cdot ]$ and $\{ \cdot, \cdot \}$ stand for the commutator and anticommutator of the involved operators, respectively.

The structure of $H_{\operatorname{sw}}$ resembles the expression of the Hamiltonian for the Swanson oscillator~\cite{Swa04,Bag15b,Fri16,Lui20}. Nevertheless, subtle but relevant differences must be noted since the coefficients $\alpha_j(t)$, $\beta_j(t)$, $j=1,2$, and $\theta(t)$, are time-dependent functions introduced to design the profile of $H_{\operatorname{sw}}$ in Eq.~(\ref{sw1}). Indeed, we distinguish four important configurations

\begin{itemize}
\item[(I)] {\sc Harmonic Oscillator.} Making $\alpha_j = \beta_j =0$ and $\theta = \frac12$, the operator (\ref{sw1}) is reduced to the well known Hamiltonian of the harmonic oscillator
\begin{equation}
\tfrac{1}{\hbar w_{0}}  H_{\operatorname{osc}}= \tfrac12 \{ \hat{a},\hat{a}^{\dagger} \}= \hat{a}^{\dagger}\hat{a}+\tfrac{1}{2}.
\label{so}
\end{equation}

\item[(II)] {\sc Hermitian Configuration.} For $\theta \in \mathbb R$, and $\alpha_j, \beta_j \in \mathbb C$ such that $\beta_j = \alpha_j^*$, with $z^{*}$ the complex-conjugate of $z \in \mathbb C$, the operator $H_{\operatorname{sw}}$ is Hermitian. The harmonic oscillator Hamiltonian (\ref{so}) is properly included in this class.

\item[(III)] {\sc Global non-Hermitian Configuration.} In general, for arbitrary complex-valued functions $\alpha_j$, $\beta_j$, and $\theta$, the operator (\ref{sw1}) is non-Hermitian. The two cases mentioned above are therefore relevant subclasses of this one. 

\item[(IV)] {\sc Non-Hermitian Configuration.} A subset of the global non-Hermitian class is characterized by real coefficients $\alpha_j$, $\beta_j$, $\theta$. This includes the harmonic oscillator as well as a subset of the Hermitian classes.
\end{itemize}

\noindent
We are interested in solving the Schr\"odinger equation defined by $H_{\operatorname{sw}}$. In the most general situation, no orthogonality is granted a priori for the solutions since both, time-dependence and non-Hermiticity, put the problem out of the Sturm-Liouville formalism. In the same context, the boson operators $\hat a$ and $\hat a^{\dagger}$ are not necessarily the ladder operators for the corresponding set of solutions. 

Without loss of generality we shall consider the Hermitian and non-Hermitian classes, as they are described above. The analysis of the global non-Hermitian case will be provided elsewhere, as it requires a more elaborated treatment.

The operator $H_{\operatorname{sw}}$ may be expressed in terms of the quadratures of position and momentum. For if one uses the well known relationships
\begin{equation}
\hat a = \sqrt{ \tfrac{m_0 w_0}{2 \hbar} } \left( \hat x + \frac{i}{m_0 w_0} \hat p_x \right), \quad 
\hat a^{\dagger} = \sqrt{ \tfrac{m_0 w_0}{2 \hbar} } \left( \hat x - \frac{i}{m_0 w_0} \hat p_x \right),
\end{equation}
then (\ref{sw1}) acquires the quadratic form
\begin{equation}
H_{\operatorname{sw}}(t)=\frac{\hat{p}_{x}^{2}}{2m(t)}+\frac{m(t)w^{2}(t)}{2}\hat{x}^{2}+i\Omega(t)\{\hat{x},\hat{p}_{x}\}+i v(t)\hat{p}_{x}+ F(t)\hat{x} \, ,
\label{Hsw2}
\end{equation}
where the linear terms are characterized by the time-dependent functions
\begin{equation}
\Omega(t):= - w_{0}\left[ \alpha_{2}(t) - \beta_{2}(t) \right] \, , \quad 
v(t):=- \sqrt{\tfrac{\hbar w_{0}}{2m_{0}}} \left[ \alpha_{1}(t) - \beta_{1}(t) \right] \, ,
\label{para2}
\end{equation}
and
\begin{equation}
F(t):=\sqrt{\tfrac{m_{0}\hbar w_{0}^{2}}{2}}\left[ \alpha_{1}(t)+\beta_{1}(t) \right].
\label{F}
\end{equation}
The mass and frequency terms of (\ref{Hsw2}) are also time-dependent
\begin{equation}
m(t)=\frac{m_{0}}{2 \theta(t)- \left[ \alpha_{2}(t) + \beta_{2}(t) \right] } \, , \quad w^{2}(t)=w_{0}^{2}\left( 4\theta^2(t)- \left[ \alpha_{2}(t)+\beta_{2}(t) \right]^{2} \right) \, .
\label{para1}
\end{equation}
Note that the above formulae are defined by the expressions $\alpha_j \pm \beta_j$, which facilitates their identification according with the classification provided above. Next we provide explicit expressions. 

Before proceeding one may revert the previous relationships, and write down the parameters of the bosonic representation~\eqref{sw1} in terms of those of the quadrature representation~\eqref{Hsw2}. We thus get
\begin{equation}
\begin{aligned}
& \alpha_{2}(t)=\frac{m^{2}(t)w^{2}(t)-m_{0}^{2}w_{0}^{2}-2m_{0}w_{0}m(t)\Omega(t)}{4m_{0}w_{0}^{2}m(t)} \, , \\[2ex]
& \beta_{2}(t)=\frac{m^{2}(t)w^{2}(t)-m_{0}^{2}w_{0}^{2}+2m_{0}w_{0}m(t)\Omega(t)}{4m_{0}w_{0}^{2}m(t)} \, ,
\end{aligned}
\label{para3}
\end{equation}
together with 
\begin{equation}
\begin{aligned}
& \alpha_{1}(t)=\frac{-m_{0}\sqrt{w_{0}}v(t)+F(t)}{\sqrt{2\hbar m_{0}w_{0}^{2}}} \, , \quad 
\beta_{1}(t)=\frac{m_{0}\sqrt{w_{0}}v(t)+F(t)}{\sqrt{2\hbar m_{0}w_{0}^{2}}} \, , \\[2ex]
& \theta(t)=\frac{m_{0}^{2}w_{0}^{2}+m^{2}(t)w^{2}(t)}{4m_{0}w_{0}^{2}m(t)}.
\end{aligned}
\label{para4}
\end{equation}

$\bullet$ {\bf Hermitian configuration.} For $\beta_j = \alpha_j^*$ the system (\ref{para2})--(\ref{F}) gives
\begin{equation}
i\Omega = \Omega_I:= 2 w_{0} \operatorname{Im} (\alpha_2), \quad iv  =v_I:= \sqrt{\tfrac{2 \hbar w_{0}}{m_{0}}}  \operatorname{Im} (\alpha_1), \quad F_R =  \sqrt{ 2 m_{0}\hbar w_{0}^{2}}  \operatorname{Re} (\alpha_1),
\end{equation}
together with
\begin{equation}
m_R= \frac{m_0}{2 \theta -2 \operatorname{Re} (\alpha_2)}, \quad w_R^2 = 4 w_0^2 \left( \theta^2 + \left[ \operatorname{Re}(\alpha_2) \right]^2 \right).
\end{equation}
Then (\ref{Hsw2}) is written in the self-adjoint form
\begin{equation}
\widetilde H_{\operatorname{sw}}(t)=\frac{\hat{p}_{x}^{2}}{2m_R(t)}+\frac{m_R(t) w_R^{2}(t)}{2}\hat{x}^{2}+ \Omega_I (t)\{\hat{x},\hat{p}_{x}\}+ v_I (t)\hat{p}_{x}+ F_R(t)\hat{x} \, .
\end{equation}
Clearly, there is a one-to-one correspondence between the sets $\mathcal{S}_{bos}=\{\alpha_{1}, \alpha_{2}, \beta_{1}, \beta_{2},\theta\}$ and $\mathcal{S}_{quad}=\{m, w,\Omega, v, F\}$. Thus, by fixing the parameters in $\mathcal{S}_{bos}$, one can determine the parameters in $\mathcal{S}_{quad}$, and vice versa.

$\bullet$ {\bf Non-Hermitian configuration.} For real coefficients $\alpha_j$, $\beta_j$ and $\theta$, the equations (\ref{Hsw2})--(\ref{para1}) yield $H_{\operatorname{sw}}$ such that $H_{\operatorname{sw}}^{\dagger} \neq H_{\operatorname{sw}}$. The non-Hermiticity is due to the real-valued functions $\Omega$ and $v$, which may be cancelled by making $\alpha_j = \beta_j$. Noticeably, the latter case is consistent with the Hermitian configuration mentioned in the previous item after making $\Omega_I=v_I=0$. For $\Omega_I \neq 0$ and $v_I \neq 0$, the appropriate transformation shows that, providing $\Omega=v=0$, the self-adjoint operator $\widetilde H_{\operatorname{sw}}$ coincides with $H_{\operatorname{sw}}$. 

Considering the above remarks, the model will be developed within the non-Hermitian configuration defined by real coefficients $\alpha_j$, $\beta_j$, $\theta$. The Hermitian configuration will be recovered after making $\alpha_j = \beta_j$. On the other hand, without loss of generality, hereafter we make $F=0$.  From (\ref{F}), the latter implies $\beta_1 = -\alpha_1$, so that $v = \sqrt{\tfrac{2\hbar w_{0}}{m_{0}}} \beta_1$.

\subsection{Space of solutions}

In position representation, the solutions of the Schr\"odinger equation defined by the Hamiltonian of the harmonic oscillator (\ref{so}) are well known to be
\begin{equation}
\Psi_n(y, \tau) = e^{-iE_n \tau } \Phi_n(y), \quad E_{n}=\hbar w_{0}(n+ \tfrac12 ) \, , \quad n=0,1,\ldots,
\label{misol}
\end{equation}
where $\tau$ stands for the time-variable, and the normalized functions
\begin{equation}
\Phi_{n}(y)=\sqrt{\frac{1}{2^{n}n!}\sqrt{\frac{m_{0}w_{0}}{\pi\hbar}}} \, e^{-\frac{m_{0}w_{0}}{2\hbar}y^{2}}H_{n}\left(\sqrt{\frac{m_{0}w_{0}}{\hbar}} y \right)\, , \quad 
\label{sol-osc-1}
\end{equation}
satisfy the eigenvalue equation $H_{\operatorname{osc}} \Phi_n (y)= E_n \Phi_n(y)$, with $H_{n}(z)$ the Hermite polynomials~\cite{Olv10}. 

To construct the solutions of the Schr\"odinger equation defined by $H_{\operatorname{sw}}$, we use the approach introduced in Ref.~\cite{Zel20c}, which is based on the formalism of point transformations. Detailed information is provided in Appendix~\ref{ApA}. Using $t$ and $x$ for the time-variable and position-coordinate of the system governed by $H_{\operatorname{sw}}$, the transformation
\begin{equation}
\tau (t)=\int^{t}\frac{dt'}{\sigma^{2}(t')}, \quad y(x,t)=\frac{\mu(t)x+\gamma(t)}{\sigma(t)}, \quad m(t)=m_{0}\mu^{2}(t) \, ,
\label{TauY}
\end{equation}
permits us to express the solutions we are looking for in terms of the formulae (\ref{misol})-(\ref{sol-osc-1}). The time-dependent functions $\gamma(t)$ and $\sigma(t)$ are to be determined.

Within the point transformation approach, the rules (\ref{TauY}) lead to a time-dependent potential in the $(x,t)$-configuration. Explicitly
\begin{multline}
V(x,t) = \frac{m_{0}\mu^{2}}{2}\left(\frac{\dot{\mathcal{W}}_{\mu}}{\mu\sigma}+\frac{2i}{\mu^{2}}\frac{d}{dt}(\mu^{2}\Omega)-4\Omega^{2}+\frac{w_{0}^{2}}{\sigma^{4}}\right)x^{2} + \\
m_{0}\mu\left( \frac{\dot{\mathcal{W}}_{\gamma}}{\sigma}-2i\mu\Omega v + \frac{i}{\mu}\frac{d}{dt}\left(\mu^{2} v\right)+\frac{w_{0}^{2}\gamma}{\sigma^{4}} \right)x+V_{0}(t) \, ,
\label{V2}
\end{multline}
with $\mathcal{W}_{\mu}$ and $\mathcal{W}_{\gamma}$ two functions of time defined in Eq.~(\ref{miW}), and $V_0(t)$ given in Eq.~(\ref{miV2}) of Appendix~\ref{ApA}. 

In turn, the wave-functions $\psi_n(x,t)$ are given by
\begin{equation}
\psi_{n}(x,t)=e^{-iE_{n}\tau(t)/\hbar}\phi_{n}(x,t) \, , 
\label{psi3}
\end{equation}
where the functions $\phi_n(x,t)$ are constructed through the eigenfunctions of the stationary  $(y,\tau)$-system:
\begin{equation}
\phi_{n}(x,t) = A_{0}^{-1}(x,t)  A_{1}^{-1}(x,t)\Phi_{n}(y(x,t)) \, ,
\label{mipsi3}
\end{equation}
with 
\begin{equation}
A_{0}(x,t) =\exp\left[ i\tfrac{m_{0}}{\hbar}\tfrac{\mu}{\sigma}\left(\tfrac{\mathcal{W}_{\mu}}{2}x^{2}+\mathcal{W}_{\gamma}x+\xi_{1}\right)\right]  \, ,
\label{miA2}
\end{equation} 
\begin{equation}
A_{1}(x,t):=\exp\left[ -\frac{m_{0}}{\hbar}\mu^{2}\left( \Omega x^{2}+v x + \xi_{2} \right) \right] \,  ,
\label{A2}
\end{equation}
and
\begin{equation}
\frac{\mu}{\sigma}\xi_{1}=\frac{\gamma W_{\gamma}}{2\sigma}+\int^{t}dt' \left( \frac{\mu^{2}(t')v^{2}(t')}{2}-\mu(t')v(t')\Omega(t')\gamma(t')\right).
\end{equation}
Thus, providing the solutions (\ref{misol}) and (\ref{sol-osc-1}) of the stationary $(y, \tau)$-system, we automatically obtain the solutions (\ref{psi3})-(\ref{A2}) of a time-dependent system in the $(x,t)$-configuration.

Note that we have conveniently introduced the form of $\psi$ given in~\eqref{psi3} so that the time-dependent factor $e^{-iE_{n}\tau(t)/\hbar}$ can be immediately identified with the phase introduced by Lewis-Riesenfeld~\cite{Lew69} for the parametric oscillator. In this form, the point transformation provides a straightforward mechanism to determine such a factor, which is nothing but the transformation of the unitary time-evolution phase of the stationary oscillator. Alternative approaches for non-Hermitian Hamiltonians have been previously studied in~\cite{Lui20}.

We would also like to emphasize that the expression of $V(x,t)$ introduced in Eq.~(\ref{V2}) represents a wide resource of complex-valued, time-dependent, potentials linked to the stationary harmonic oscillator. The applicability of the above results is therefore very wide. This embraces time-dependent Hermitian oscillators for real-valued functions $V(x,t)$ as well as non-Hermitian oscillators (stationary and nonstationary) for complex-valued functions $V(x,t)$. The main point is the manipulability of the concrete form of $V(x,t)$ by tuning its time-dependent coefficients.

In the present work we concentrate in the relationship between $V(x,t)$ and the potential part of $H_{\operatorname{sw}}$, written in coordinate representation (\ref{Hsw2}). Other oscillators will be studied elsewhere.

\subsection{Time-dependent model with non-Hermiticity}

Comparing the non-kinetic part of $H_{\operatorname{sw}}$ with $V(x,t)$ leads to Eqs.~(\ref{eqs1})-(\ref{miseqs1}) of Appendix~\ref{ApA}. The time-dependent functions $\gamma$ and $\sigma$ that define the transformation (\ref{TauY}) are solutions of (\ref{eqs1})  and (\ref{misAeqs1}), respectively. These equations include pure-imaginary terms that may be canceled through the constraints
\begin{equation}
\mu^2 \Omega  =\Omega_{0} \, , \quad \mu^2 v =v_{0} \, ,
\label{micond}
\end{equation}
where $\Omega_0$ and $v_0$ are real  constants to be fixed. The functions $\gamma$ and $\sigma$ are therefore defined by the following system of equations
\begin{equation}
\ddot{\sigma}+\left(w^{2}+4\frac{\Omega^{2}_{0}}{\mu^{4}}-\frac{\ddot{\mu}}{\mu} \right)\sigma=\frac{w_{0}^{2}}{\sigma^{3}}, \quad
\ddot{\gamma} + \left(w^{2}+4\frac{\Omega^{2}_{0}}{\mu^{4}}-\frac{\ddot{\mu}}{\mu} \right) \gamma = 2\frac{v_{0}\Omega_{0}}{\mu^{3}},
\label{miseqs2}
\end{equation}
together with
\begin{equation}
\frac{d}{dt}\left(\frac{\mu}{\sigma}\xi_{1}+i\mu^{2}\xi_{2}-\frac{\gamma}{2\sigma} \mathcal{W}_{\gamma} -i\frac{\hbar}{2m_{0}}\ln\frac{\mu}{\sigma} \right) - \frac{v^{2}_{0}}{2\mu^{2}}+v_{0}\Omega_{0}\frac{\gamma}{\mu^{3}}-\frac{\hbar}{m_{0}}\frac{\Omega_{0}}{\mu^{2}}=0 \, .
\label{eqs2}
\end{equation}
The real-valued functions $\xi_{1}(t)$ and $\xi_{2}(t)$ resulted from integration with respect to $x$. Note that the equation for $\sigma$ in (\ref{miseqs2}) has the structure of the nonlinear differential equation named after Ermakov~\cite{Erm08}. Detailed information about the method of solution and applications can be consulted in, e.g., \cite{Ros15,Bla19,Sch18}.

On the other hand, combining (\ref{micond}) with (\ref{para3}) and (\ref{para4}) gives rise to the set of parameters $\mathcal{S}_{bos}$ of the bosonic representation~\eqref{sw1}. Therefore, the operator introduced in Eq.~(\ref{Hsw2}) acquires a simpler form
\begin{equation}
H_{\operatorname{sw}}(t) =\frac{\hat{p}_{x}^{2}}{2m_{0}\mu^{2}(t)}+\frac{m_{0}\mu^{2}(t)w^{2}(t)}{2}\hat{x}^{2}+\frac{i}{\mu^{2}(t)}\left[ \Omega_{0}\{\hat{x},\hat{p}_{x}\}+v_{0}\hat{p}_{x} \right] \, ,
\label{Hsw3}
\end{equation}
where $\mu$ plays the role of a time-dependent mass.

The non-Hermiticity of $H_{\operatorname{sw}}$ in Eq.~(\ref{Hsw3}) is parameterized by the real constants $\Omega_0$ and $v_0$; the Hermitian version of this operator arises by turning off both of these parameters. Paying attention to the transformation of the position-variable (\ref{TauY}), we realize that $y(x,t)$ is real-valued if the $x$-coordinate is real, as expected. The construction of non-Hermitian operators $H_{\operatorname{sw}}$ permitting complex-valued mappings for $y(x,t)$ may be derived from a more general scheme, which is out of the scope of the present work.

Hereafter we take $\Omega_0 \geq 0$ and $v_0 \geq 0$. The complete characterization of the operator (\ref{Hsw3}) is therefore provided by the analytical form of the time-dependent functions $\alpha_2$ and $\beta_2$. Remarkably, up to the non-Hermitian term, the expression of  $H_{\operatorname{sw}}$ in (\ref{Hsw3}) has the structure of the generalized Caldirola-Kanai oscillator discussed in \cite{Ald11} within the Arnold transformation approach. Here, the operator (\ref{Hsw3}) is linked with a time-dependent oscillator in the non-Hermitian regime for which both, mass and frequency, depend on time. As the Hermitian limit $\Omega_{0}=v_{0}=0$ has already been treated in~\cite{Ald11,Zel21b}, it will be discarded throughout the rest of the manuscript.

\section{Applications}
\label{applica}
To illustrate the applicability of our approach we consider a model generated by the time-dependent functions
\begin{equation}
\mu^2= e^{-\Gamma t}, \quad w^2=w_0^2.
\label{misf}
\end{equation}
The formulae (\ref{para3})-(\ref{para4}) give
\begin{equation}
\begin{aligned}
&\alpha_2 (t) = -\tfrac12 \sinh (\Gamma t) - \frac{\Omega_{0}}{2w_{0}}e^{\Gamma t} , \quad \beta_2 (t) = -\tfrac12 \sinh (\Gamma t) + \frac{\Omega_{0}}{2w_{0}}e^{\Gamma t} , \\ 
&\theta = \tfrac12 \cosh (\Gamma t), \quad \beta_1 = \sqrt{ \tfrac{m_0}{2\hbar w_0}} v_0 e^{\Gamma t}.
\end{aligned}
\end{equation}
Therefore, the operator (\ref{Hsw3}) acquires the form
\begin{equation}
H_{\operatorname{sw}}(t)=\frac{e^{\Gamma t}}{2m_{0}} \hat{p}_{x}^{2} +\frac{m_{0}w_{0}^{2}}{2} e^{-\Gamma t} \hat{x}^{2}+ie^{\Gamma t}\left( \Omega_{0}\{\hat{x},\hat{p}_{x}\}+v_{0}\hat{p}_{x} \right) \, .
\label{CK-nH}
\end{equation}
The self-adjoint part of this operator coincides with the Hamiltonian studied independently by Caldirola \cite{Cal41} and Kanai \cite{Kan48}. That is, the Hamiltonian $H_{\operatorname{sw}}(t)$ introduced in Eq.~(\ref{CK-nH}) may be considered a non-Hermitian extension of the  Caldirola-Kanai oscillator.

At the classical level, the Caldirola-Kanai Hamiltonian leads to the Newton equation of motion, including a friction term that is proportional to velocity. This property motivated the unfinished debate about the nature and possible interpretation of friction forces in quantum mechanics~\cite{Gre79}. Admitted as a very interesting problem in the formal structure of quantum mechanics, the Caldirola-Kanai oscillator deserves particular attention. For instance, it has been studied in terms of the quantum Arnold transformation\cite{Ald11,Gue15b}, where exact solutions have been provided for the related Schr\"odinger equation.

Considering the interest that non-Hermitian structures like the Hamiltonian (\ref{CK-nH}) arouse in the literature on the matter, we solve the corresponding dynamical law in both the classical and quantum pictures.

\subsection{Classical picture} 
\label{sec:class-sol}

To determine the classical equations of motion associated to the classical counterpart of operator~\eqref{CK-nH}, consider the classical Hamiltonian 
\begin{equation}
H_{\operatorname{class}}(Q,P;t)=\frac{e^{\Gamma t}P^{2}}{2m_{0}}+\frac{e^{-\Gamma t}m_{0}w_{0}^{2}}{2}Q^{2}+ie^{\Gamma t}\left(2\Omega_{0}QP+v_{0}P \right) \, .
\label{CK-class}
\end{equation}
Using canonical quantization, together with the symmetrization rule $QP\rightarrow \frac{1}{2}\left\{ \hat{x},\hat{p}_{x} \right\}$, operator~\eqref{CK-nH} is recovered, as expected.

The Hamilton equations of motion, $\dot{Q}=\frac{\partial H_{\operatorname{class}}}{\partial P}$ and $-\dot{P}=\frac{\partial H_{\operatorname{class}}}{\partial Q}$, yield 
\begin{equation}
P=m_{0}e^{-\Gamma t}\dot{Q}-2im_{0}(\Omega_{0}Q+v_{0}) \, ,
\label{CanP}
\end{equation}
which is complex-valued in the configuration space $( Q,\dot{Q} )$. From~\eqref{CanP} we have $\dot{Q}=e^{\Gamma t}(m_{0}^{-1}P+2i(\Omega_{0}Q+v_{0}))$, so that $\ddot{Q}=\{\dot{Q},H_{\operatorname{class}}\}_{\operatorname{PB}}+\frac{\partial \dot{Q}}{\partial t}$, where $\{\cdot,\cdot\}_{\operatorname{PB}}$ stands for the Poisson bracket. After some calculations one arrives at equation
\begin{equation}
\ddot{Q}-\Gamma \dot{Q}+\left( w_{0}^{2}+4\Omega_{0}^{2}e^{2\Gamma t} \right)Q=-2\Omega_{0}v_{0}e^{2\Gamma t} \, ,
\label{class-q1}
\end{equation}
which defines the behavior of a damped parametric oscillator with time-dependent frequency $(w_{0}^{2}+4\Omega_{0}^{2}e^{2\Gamma t})$, and subjected to the force $-2\Omega_{0}v_{0}e^{2\Gamma t}$. 

The second-order differential equation with real-valued coefficients~\eqref{class-q1} admits real-valued solutions $Q(t)$ upon appropriate initial conditions $Q(0),\dot{Q}(0)\in\mathbb{R}$. To write it in a more familiar form we make $\mathcal{Q}=e^{-\Gamma t/2}Q$, a change of variable known as \textit{expanding coordinates}~\cite{Ped87,Sch90}. Then
\begin{equation}
\ddot{\mathcal{Q}}+\left( w_{0}^{2}-\frac{\Gamma^{2}}{4}+4\Omega_{0}^{2}e^{2\Gamma t} \right)\mathcal{Q}=-2\Omega_{0}v_{0}e^{3\Gamma t/2} 
\label{class-q2}
\end{equation}
is the equation of motion for a driven parametric oscillator with no damping. 

\begin{figure}
\centering
\subfloat[][]{\includegraphics[width=0.3\textwidth]{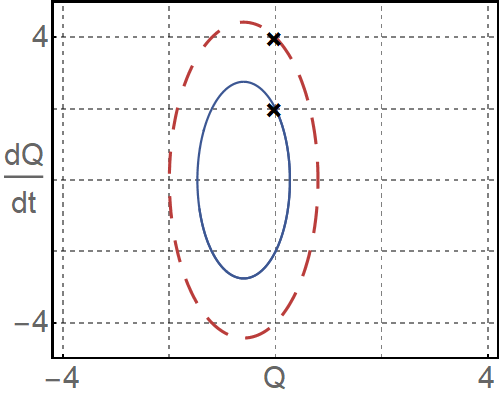}
\label{F1a}}
\hspace{2mm}
\subfloat[][]{\includegraphics[width=0.37\textwidth]{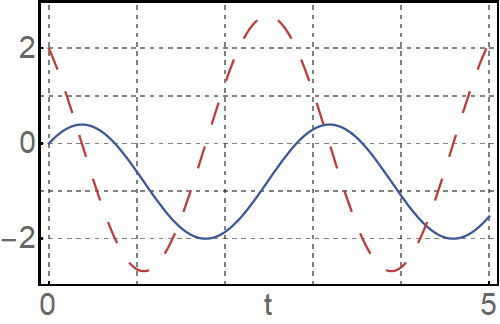}
\label{F1b}}
\caption{(a) Trajectories in the configuration space $(\mathcal{Q},\dot{\mathcal{Q}})$ for constant-mass ($\Gamma=0$), with $w_{0}=1$, $\Omega_{0}=1.5$, and $v_{0}=2$. The initial conditions are $\mathcal{Q}(0)=0$, $\dot{\mathcal{Q}}(0)=2$ (blue-solid), and $\mathcal{Q}(0)=0$, $\dot{\mathcal{Q}}(0)=4$ (red-dashed). The cross indicates the initial conditions. (b) The time-dependent functions $\mathcal{Q}$ (blue-solid) and $\dot{\mathcal{Q}}$ (red-dashed) for the initial conditions $\mathcal{Q}(0)=0$, $\dot{\mathcal{Q}}(0)=2$.
}
\label{F1}
\end{figure}

$\bullet$ {\bf Constant mass.} For $\Gamma=0$ the mass term in (\ref{CK-class}) is $m_0 = \operatorname{const}$. In this case the solutions of Eq.~(\ref{class-q2}) are immediate
\begin{equation}
\begin{aligned}
\mathcal{Q}(t)=\mathcal{A}\cos\left(\sqrt{w_{0}^{2}+4\Omega_{0}^{2}}\, t+\varphi\right)-\frac{2\Omega_{0}v_{0}}{w_{0}^{2}+4\Omega_{0}^{2}} \, , \\
\dot{\mathcal{Q}}(t)=-\mathcal{A}\sqrt{w_{0}^{2}+4\Omega_{0}^{2}}\sin\left(\sqrt{w_{0}^{2}+4\Omega_{0}^{2}}\,t+\varphi\right) \, ,
\end{aligned}
\label{const-mass}
\end{equation}
with $\mathcal{A}$ and $\varphi$ integration constants determined from the initial conditions as
\begin{equation}
\mathcal{A}=
\sqrt{\dot{\mathcal{Q}}^{2}(0)+\left( \mathcal{Q}(0) +\tfrac{2\Omega_{0}v_{0}}{w_{0}^{2}+4\Omega_{0}^{2}} \right)^{2}}
 \, , \quad \tan\varphi=-\tfrac{\sqrt{w_{0}+4\Omega_{0}^{2}}\dot{\mathcal{Q}}(0)}{2\Omega_{0}v_{0}+(w_{0}^{2}+4\Omega_{0}^{2})\mathcal{Q}(0)} \, .
\end{equation}
The dynamics on the configuration space $(\mathcal{Q},\dot{\mathcal{Q}})$ describes elliptic closed trajectories, displaced along the $Q$ axis due to the presence of both $\Omega_{0}$ and $v_{0}$, see Figure~\ref{F1}. 

$\bullet$ {\bf Time-dependent mass.} For $\Gamma \neq 0$ we may introduce the variable $z=e^{\Gamma t}$ to get
\begin{equation}
z^{2}\frac{d^{2}\mathcal{Q}}{dz^{2}}+z\frac{d\mathcal{Q}}{dz}+\left(\Lambda^{2}+4\overline{\Omega_{0}}^{2}z^{2} \right)\mathcal{Q}=-2\overline{v}_{0}\overline{\Omega}_{0}z^{3/2} \, ,
\label{class-q3}
\end{equation}
where
\begin{equation}
\Lambda^{2}:=\frac{w_{0}^{2}}{\Gamma^{2}}-\frac{1}{4} \, , \quad \overline{\Omega}_{0}=\frac{\Omega_{0}}{\Gamma} \, , \quad \overline{v}_{0}=\frac{v_{0}}{\Gamma} \, .
\label{class-q4}
\end{equation}
It is immediate to identify that the homogeneous part of (\ref{class-q3}) coincides with the Bessel differential equation \cite{Olv10}. In this case we use the solutions
\begin{equation}
\mathcal{Q}_{h;1}(t)=\frac{J_{i\Lambda}(2\overline{\Omega}_{0}e^{\Gamma t})}{\overline{\Omega}_{0}^{i\Lambda}} \, , \quad 
\mathcal{Q}_{h;2}(t)\equiv \mathcal{Q}_{h;1}^{*}(t)=\frac{J_{-i\Lambda}(2\overline{\Omega}_{0}e^{\Gamma t})}{\overline{\Omega}_{0}^{-i\Lambda}} \, ,
\label{QQ1}
\end{equation}
where $J_{\nu}(z)$ is the Bessel function of the first kind, and the constants $\overline{\Omega}_{0}^{\pm i\Lambda}$ have been introduced such that
\begin{equation}
\lim_{\overline{\Omega}_{0}\rightarrow 0}\mathcal{Q}_{h;1}(t)\rightarrow \frac{\exp[i\sqrt{w_{0}^{2}-\Gamma^{2}/4}\, t]}{\Gamma(1+i\Lambda) } \, , \quad 
\lim_{\overline{\Omega}_{0}\rightarrow 0}\mathcal{Q}_{h;2}(t)\rightarrow \frac{\exp[-i\sqrt{w_{0}^{2}-\Gamma^{2}/4}\, t]}{\Gamma (1-i\Lambda)} \, ,
\end{equation} 
lead to the solutions of the conventional oscillator.

\begin{figure}
\centering
\subfloat[][]{\includegraphics[width=0.3\textwidth]{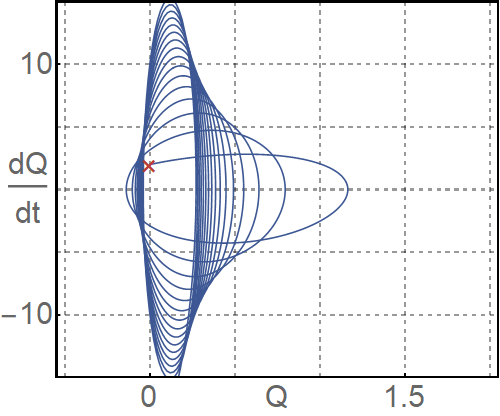}
\label{F2a}}
\hspace{2mm}
\subfloat[][]{\includegraphics[width=0.38\textwidth]{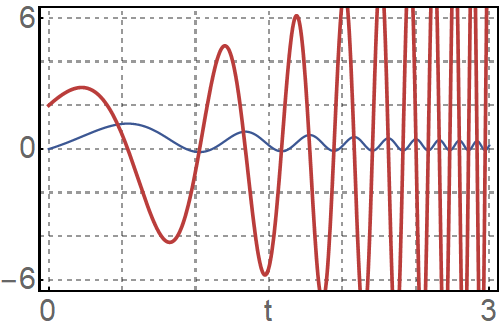}
\label{F2b}}
\caption{(a) Trajectories in the configuration space $(\mathcal{Q},\dot{\mathcal{Q}})$ for $\Gamma=1$ in the interval $t\in(0,3)$, with $w_{0}=1$, $\Omega_{0}=1.5$, and $v_{0}=2$. The initial conditions are $\mathcal{Q}(0)=0$, and $\dot{\mathcal{Q}}(0)=2$. The red cross depicts the initial condition. (b) The time-dependent functions $\mathcal{Q}(t)$ (blue-solid) and $\dot{\mathcal{Q}}(t)$ (red-thick-solid) for the above mentioned parameters.}
\label{F2}
\end{figure}

On the other hand, after some calculations the function
\begin{multline}
\mathcal{Q}_{p}(t)=i\frac{\pi\overline{v}_{0}\overline{\Omega}_{0}^{1-i\Lambda}}{2\sinh(\pi\Lambda)} \frac{\Gamma(\frac{3}{4}-i\frac{\Lambda}{2})}{\Gamma(\frac{7}{4}-i\frac{\Lambda}{2})\Gamma(1-i\Lambda)} \times \\
e^{(\frac{3}{2}-i\Lambda)\Gamma t}J_{i\Lambda}\left( 2\overline{\Omega}_{0}e^{\Gamma t} \right)
\, {}_{1}F_{2}
\left(\left.\begin{matrix}
\frac{3}{4}-i\frac{\Lambda}{2} \\
1-i\Lambda,\frac{7}{4}-i\frac{\Lambda}{2}
\end{matrix}
\right\vert -\overline{\Omega}_{0}^{2}e^{2\Gamma t} \right) \, ,
\end{multline}
provides the particular solution of equation~\eqref{class-q3}. Then, the general solution can be written in the form
\begin{equation}
\mathcal{Q}(t)=\ell_{1}\operatorname{Re}Q_{h;1}(t)+\ell_{2}\operatorname{Im}\mathcal{Q}_{h;1}(t)+2\operatorname{Re}\mathcal{Q}_{p}(t) \, ,
\label{exp-mass-gen}
\end{equation}
where $\ell_{1}$ and $\ell_{2}$ are arbitrary real constants fixed from the initial conditions. In contradistinction to the constant mass solutions, an explicit form for $\ell_{1}$ and $\ell_{2}$ in terms of the initial conditions $\mathcal{Q}(0)$ and $\dot{\mathcal{Q}}(0)$ is not feasible; however, it can be established by numerical means. 

The corresponding trajectory in the configuration space is depicted in Figure~\ref{F2a} for a finite time interval. Clearly, the trajectory is no longer closed as the particle mass is continuously changing on time. Remark that, as time pass by, the particle localizes to a well defined finite region that shrinks on time, while the velocity increases exponentially. This behavior is clear from Figure~\ref{F2b}.

\subsection{Quantum picture}

The solutions of the Schr\"odinger equation defined by the Hamiltonian~\eqref{CK-nH} require to have at hand the functions $\sigma$ and $\gamma$. In this regard, the solutions of the homogeneous equation
\begin{equation}
\ddot{q}+\left( w_{0}^{2}-\frac{\Gamma^{2}}{4}+4\Omega_{0}^{2}e^{2\Gamma t} \right)q=0 \, 
\label{q-CK1}
\end{equation}
serve to solve both, the Ermakov and the inhomogeneous equations included in \eqref{miseqs2}. Indeed, following \cite{Ros15,Bla19} we know that the Ermakov equation defining $\sigma$ is solved by using the two linearly independent solutions of (\ref{q-CK1}), namely $q_{1}$ and $q_{2}$. We thus get 
\begin{equation}
\sigma(t)=\left(a q_{1}^{2}(t)+b q_{1}(t)q_{2}(t) +c q_{2}^{2}(t)\right)^{\frac{1}{2}} \, , \quad b^{2}-4ac=-\frac{w_{0}^{2}}{W_{0}^{2}} \, ,
\end{equation}
with $W_{0}=\operatorname{Wr}(q_{1},q_{2})$ the Wronskian of $q_{1}$ and $q_{2}$, which in this case is always a constant. In turn, the inhomogeneous equation associated with $\gamma$ shares solutions with the classical equation \eqref{class-q2}, already solved in the previous section.

\begin{figure}
\centering
\subfloat[][$\Gamma=0$]{\includegraphics[width=0.3\textwidth]{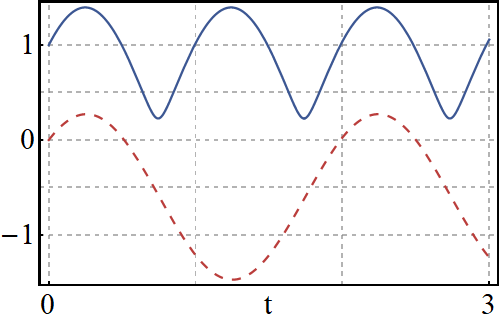}}
\hspace{3mm}
\subfloat[][$\Gamma=1$]{\includegraphics[width=0.3\textwidth]{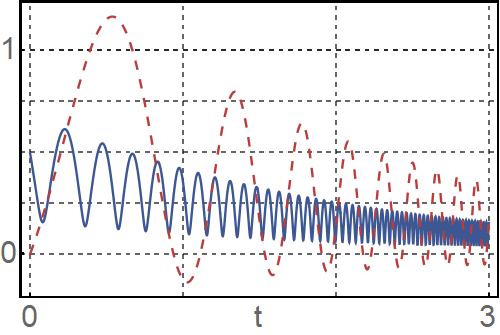}}
\caption{Solution to the Ermakov equation $\sigma(t)$ (blue-solid) and the non-homogeneous equation $\gamma(t)$ (red-dashed) for the mass-term $e^{-\Gamma t/2}$ for the constant mass case $\Gamma=0$ (a) and $\Gamma=1$ (b). Moreover, in (a) we have used $\mathcal{A}$ and $\varphi$ as in Figure~\ref{F1}, whereas in (b) we have used $\ell_{1}$ and $\ell_{2}$ as in Figure~\ref{F2}. The remaining parameters have been selected in both cases as $a=c=w_{0}=1$, $\Omega=1.5$, $v_{0}=2$.}
\label{F3}
\end{figure}

\begin{itemize}
\item For the constant mass case, we use the homogeneous solutions and their respective Wronskian $W_{0}$ as
\begin{equation}
q_{1}=\cos\left(\sqrt{w_{0}^{2}+4\Omega_{0}^{2}}\, t\right) \, , \quad 
q_{2}=\sin\left(\sqrt{w_{0}^{2}+4\Omega_{0}^{2}}\, t\right) \, , \quad W_{0}=\sqrt{w_{0}^{2}+4\Omega_{0}^{2}} \, ,
\end{equation}
whereas, the solution for $\gamma(t)$ is the same as that for $-\mathcal{Q}(t)$ provided in~\eqref{const-mass}.

\item For the mass-term $\mu(t)=e^{-\Gamma t/2}$, we have
\begin{equation}
q_{1}(t)=\operatorname{Re}\left(\frac{J_{i\Lambda}(2\overline{\Omega}_{0}e^{\Gamma t})}{\overline{\Omega}_{0}^{i\Lambda}}\right) \, , \quad 
q_{2}(t)=\operatorname{Im}\left(\frac{J_{i\Lambda}(2\overline{\Omega}_{0}e^{\Gamma t})}{\overline{\Omega}_{0}^{i\Lambda}}\right) \, , \quad W_{0}=\frac{\Gamma}{\pi}\sinh(\pi\Lambda) \, ,
\label{qq1}
\end{equation}
with $\overline{\Omega}_{0}$ and $\Lambda$ given in~\eqref{class-q4}. Moreover, $\gamma(t)=-\mathcal{Q}(t)$, with $\mathcal{Q}(t)$ given in~\eqref{exp-mass-gen}.

\end{itemize}

The profile of $\sigma(t)$ and $\gamma(t)$ is depicted in Figure~\ref{F3} paying special attention to the constant mass $\Gamma=0$ and $\Gamma=1$ cases. In such a figure it is verified that indeed the solution to the Ermakov equation is always different to zero, as we stated earlier and proved in~\cite{Ros18,Bla19} (see also~\cite{Bar06}). Therefore, the point transformation is non-singular for $t\in\mathbb{R}$. 

\subsection{Hermitian conjugate and bi-orthogonality}

As a byproduct of the point transformation, the construction of the Hermitian conjugate  $H_{\operatorname{sw}}^{\dagger}$ and its wave-functions is immediate by noticing that
\begin{equation}
H^{\dagger}_{\operatorname{sw}}=\frac{\hat{p}_{x}^{2}}{2m_{0}\mu^{2}(t)}+m_{0}\mu^{2}(t)w^{2}(t)\hat{x}^{2}-\frac{i}{\mu^{2}(t)}\left( \Omega_{0}\{\hat{x},\hat{p}_{x}\}+v_{0}\hat{p}_{x} \right)\not=H_{\operatorname{sw}} \, 
\label{Hswd}
\end{equation}
arises from $H_{\operatorname{sw}}$ through $\Omega_{0}\rightarrow-\Omega_{0}$ and $v_{0}\rightarrow -v_{0}$. This change leaves invariant the differential equations defining $\sigma$ and $\gamma$ in~\eqref{eqs2}. Therefore, the transformed coordinate $y(x,t)$ and time parameter $\tau(t)$ are the same for both, $H_{\operatorname{sw}}$ and $H_{\operatorname{sw}}^{\dagger}$. In this form, the solutions $\widetilde{\psi}(x,t)$ of the Schr\"odinger equation associated with~\eqref{Hswd} are also obtained from the stationary solutions
\begin{equation}
\widetilde{\psi}(x,y)=(\widetilde{A}(x,t))^{-1}\Psi(y(x,t),\tau(t)) \, , \quad \widetilde{A}(x,t)=\sqrt{\frac{\sigma}{\mu}}A_{0}(x,t)(\mathcal{A}_{1}(x,t))^{-1} \, .
\label{psid1}
\end{equation}
That is,
\begin{equation}
\widetilde{\psi}_{n}(x,t)=e^{-iE_{n}\tau(t)/\hbar}\widetilde{\phi}_{n}(x,t)\, , \quad \widetilde{\phi}_{n}(x,t):=\sqrt{\frac{\mu}{\sigma}}(A_{0}(x,t))^{-1}\mathcal{A}_{1}(x,t)\Phi_{n}(y(x,t)) \, ,
\label{psid2}
\end{equation}
with 
\begin{equation}
\Phi_{n}(y(x,t))=\sqrt{\frac{1}{2^{n}n!}\sqrt{\frac{m_{0}w_{0}}{\pi\hbar}}}e^{-\frac{m_{0}w_{0}}{2\hbar}\left(\frac{\mu x+\gamma}{\sigma}\right)^{2}}H_{n}\left(\sqrt{\frac{m_{0}w_{0}}{\hbar}}\left(\frac{\mu x+\gamma}{\sigma}\right)\right) \, .
\label{Phi1}
\end{equation}

\begin{figure}
\centering
\subfloat[][$n=0$]{\includegraphics[width=0.2\textwidth]{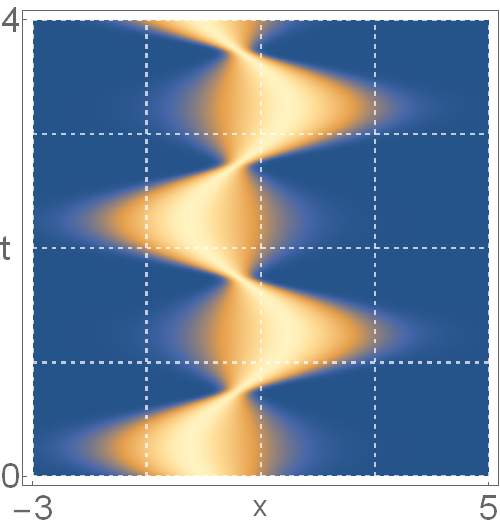}}
\hspace{3mm}
\subfloat[][$n=1$]{\includegraphics[width=0.2\textwidth]{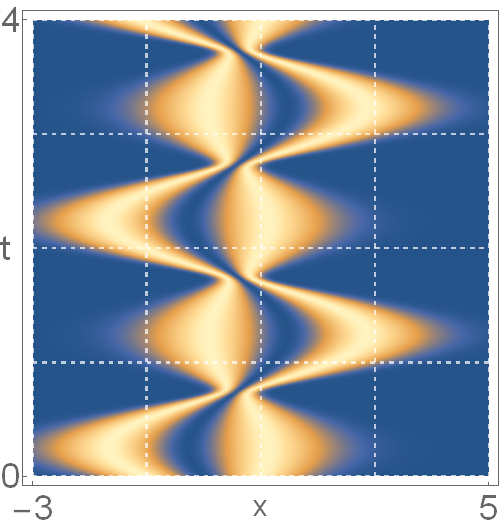}}
\hspace{3mm}
\subfloat[][$n=2$]{\includegraphics[width=0.2\textwidth]{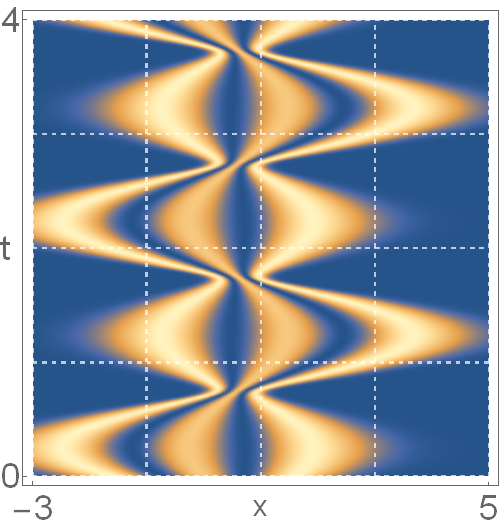}}
\\
\vspace{3mm}
\subfloat[][$n=0$]{\includegraphics[width=0.2\textwidth]{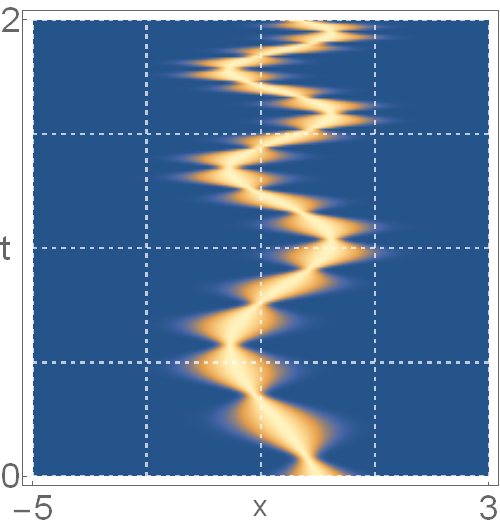}}
\hspace{3mm}
\subfloat[][$n=1$]{\includegraphics[width=0.2\textwidth]{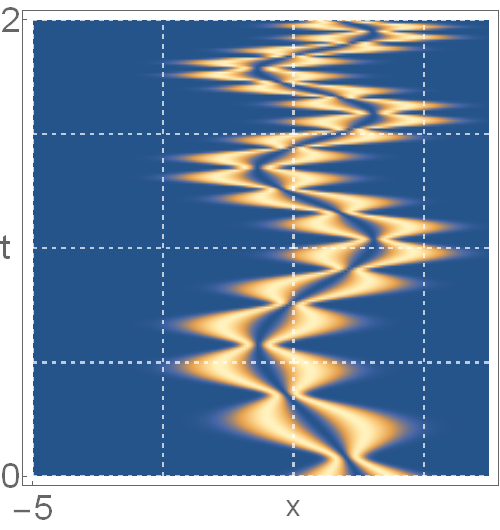}}
\hspace{3mm}
\subfloat[][$n=2$]{\includegraphics[width=0.2\textwidth]{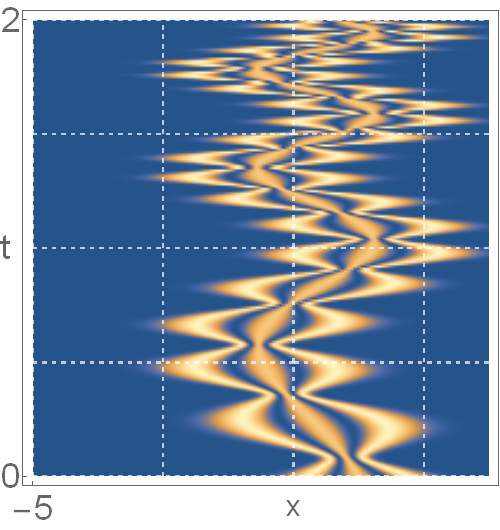}}
\caption{Probability distribution in the bi-product sense~\eqref{PD} for the constant mass case $\Gamma=0$ (first row) and the time-dependent mass case $\Gamma=1$ (second row). The parameters are the same as in Figure~\ref{F3}.}
\label{WF1}
\end{figure}

The above results are remarkable since the wave-functions $\psi_n(x,t)$, together with its Hermitian conjugate counterparts $\widetilde\psi_n(x,t)$ form a bi-orthogonal system that provides a mathematical procedure to satisfy the superposition principle \cite{Ros18}. In this form the non-Hermitian oscillators introduced above can be studied in much the same way as in the Hermitian approaches. In the present case, the probability density may be studied through the following equivalent forms~\cite{Zel20a}:
\begin{equation}
\mathcal{P}_{n}^{(B)}(x,t):=\vert\widetilde{\psi}_{n}^{*}(x,t)\psi_{n}(x,t)\vert=\vert\psi_{n}^{*}(x,t)\widetilde{\psi}_{n}(x,t)\vert \, . 
\label{PD}
\end{equation}

Figure~\ref{WF1} illustrates the time-evolution of the probability density (\ref{PD}) for the non-Hermitian oscillators introduced in the previous section. In particular, for constant mass, the probability density oscillates periodically with time, see the first row of Figure~\ref{WF1}. Note  also that the wave-packet width oscillates as well, producing a ``breathing'' effect as the wave-packet propagates. This behavior matches well with the dynamics of the classical counterpart. For $\Gamma=1$, shown in the second row of Figure~\ref{WF1}, the probability density is no longer periodic. Instead, the wave-packet follows the trajectory of a damped oscillatory.

\section{Concluding remarks}
\label{conclu}

We have shown that the stationary harmonic oscillator is connected with time-dependent systems exhibiting non-Hermiticity via point transformations. Although this association includes a very wide set of such systems, we concentrated in a generalization of the Swanson oscillator that includes the Caldirola-Kanai system as particular case. We provided concrete expressions for the Hamiltonian and the corresponding exact solutions in both pictures, classical and quantum. The systems studied here transit to their Hermitian configuration at the appropriate limit of the involved parameters. We have also shown that the point transformations provide automatically the Hermitian-conjugate of the system under study as well as the corresponding solutions. This has been used to construct a bi-orthogonal system allowing the calculation of probability densities, which are dissimilar to the conventional densities in the sense that no phase-shifts producing oscillations of the norm are allowed~\cite{Zel20a}. Noticeably, the real and imaginary parts of the fundamental solutions, as well as their probability densities, behave qualitatively equal in both normalizations, although the bi-normalized values are usually larger than the conventionally normalized ones. The latter situation is reversed for superpositions of the basis elements, see~\cite{Zel20a}. 

\subsection*{Acknowledgment}
K. Zelaya acknowledges the support from the project ``Physicist on the move II'' (KINE\'O II), Czech Republic, Grant No. CZ.02.2.69/0.0/0.0/18\_053/0017163. This research has been funded by Consejo Nacional de Ciencia y Tecnolog\'ia (CONACyT), Mexico, Grant No. A1-S-24569.

\appendix
\section{Point transformations}
\label{ApA}

\renewcommand{\thesection}{A-\arabic{section}}
\setcounter{section}{0}  

\renewcommand{\theequation}{A-\arabic{equation}}
\setcounter{equation}{0}  

In position-representation, the Schr\"odinger equation defined by the harmonic oscillator Hamiltonian (\ref{so}) may be written as
\begin{equation}
i\hbar\frac{\partial\Psi}{\partial\tau}=-\frac{\hbar^{2}}{2m_{0}}\frac{\partial^{2}\Psi}{\partial y^{2}}+\frac{m_{0}w_{0}^{2}}{2}y^{2}\Psi \, .
\label{schr-osc}
\end{equation}
The solutions are provided in Eqs.~(\ref{misol})-(\ref{sol-osc-1}) of the main text. Hereafter we say that the set $\{ y, \tau, \Psi \}$ defines the frame of the harmonic oscillator (HO), which is composited by the spatial-coordinates $y$, the time-variable $\tau$, and the solutions $\Psi$ of~\eqref{schr-osc}.

Equivalently, for the operator $H_{\operatorname{sw}}(t)$ introduced in Eq.~(\ref{Hsw2}) one has
\begin{multline}
i\hbar\frac{\partial\psi}{\partial t}=-\frac{\hbar^{2}}{2m(t)}\frac{\partial^{2}\psi}{\partial x^{2}}\\
+\left( 2\hbar\Omega(t)x 
+ \hbar v(t) \right)\frac{\partial\psi}{\partial x}
+\left[ \frac{m(t)w^{2}(t)}{2}x^{2}+F(t)x+ \hbar \Omega(t) \right] \psi \, .
\label{schr-sw}
\end{multline}
The set $\{ x, t, \psi \}$ defines the frame of the time-dependent non-Hermitian oscillator (TnH-HO), integrated by the spatial-coordinates $x$, the time-variable $t$, and the solutions $\psi$ of Eq.~(\ref{schr-sw}).

Within the point transformation theory \cite{Ste93}, the mapping from HO to TnH-HO is established by the relationships
\begin{equation}
y=y(x,t), \quad  \tau =  \tau (x,t), \quad \Psi= \Psi (y(x,t), \tau (x,t))=G(x,t;\psi(x,t)).
\label{rel1}
\end{equation}
Computing the total derivatives $\frac{d\Psi}{dx}$, $\tfrac{d\Psi}{dt}$, and $\frac{d^{2}\Psi}{dx^{2}}$, one has
\begin{equation}
\frac{\partial\Psi}{\partial \tau}=G_{1}\left( x,t;\psi,\frac{\partial\psi}{\partial x},\frac{\partial\psi}{\partial t} \right) \, , \quad \frac{\partial^{2}\Psi}{\partial y^{2}}=G_{2}\left( x,t;\psi,\frac{\partial\psi}{\partial x},\frac{\partial^{2}\psi}{\partial x^{2}},\frac{\partial\psi}{\partial t} \right) \, .
\label{rel2}
\end{equation}
To avoid nonlinear terms one may introduce the conitions~\cite{Zel20c}
\begin{equation}
\Psi=G(x,t;\psi)=A(x,t)\psi \, , \quad \tau=\tau(t) \, .
\label{rel3}
\end{equation}
After some calculations, from \eqref{rel2} one arrives at
\begin{equation}
\begin{aligned}
& \Psi_{\tau} = \frac{A}{\tau_{t}}\left[ - \frac{y_{t}}{y_{x}} \psi_x +  \psi_t+\left( \frac{A_{t}}{A}-\frac{y_{t}}{y_{x}} \frac{A_{x}}{A}\right) \psi \right], \\
& \Psi_{y,y} = \frac{A}{y_{x}^{2}}\left[ \psi_{x,x}+\left( 2 \frac{A_{x}}{A} - \frac{y_{xx}}{y_{x}} \right) \psi_x + \left(\frac{A_{xx}}{A}-\frac{y_{xx} }{y_{x}}\frac{A_{x}}{A} \right)\psi \right],
\end{aligned}
\label{rel4}
\end{equation}
the subindices denote partial derivatives, $f_u = \frac{\partial f}{\partial u}$. Substituting \eqref{rel3} and~\eqref{rel4} into~\eqref{schr-osc} gives
\begin{equation}
i\hbar \psi_t+\frac{\hbar^{2}}{2m_{0}}\frac{\tau_{t}}{y_{x}^{2}} \psi_{x,x} + B(x,t) \psi_x - V(x,t)\psi=0,
\label{PTfinal}
\end{equation}
with
\begin{equation}
\begin{aligned}
& B(x,t)=-i\hbar\frac{y_{t}}{y_{x}}+\frac{\hbar^{2}}{2m_{0}}\frac{\tau_{t}}{y_{x}^{2}}\left( 2\frac{A_{x}}{A}-\frac{y_{xx}}{y_{x}} \right) ,\\
& V(x,t)=-i\hbar\left(\frac{A_{t}}{A}-\frac
{y_{t}}{y_{x}}\frac{A_{x}}{A} \right)-\frac{\hbar^{2}}{2m_{0}}\frac{\tau_{t}}{y_{x}^{2}}\left( \frac{A_{xx}}{A}-\frac{y_{xx}}{y_{x}}\frac{A_{x}}{A} \right)+\frac{\tau_{t}}{2}m_{0}w_{0}^{2} y 	^{2}(x,t).
\end{aligned}
\label{BV}
\end{equation}
Demanding coincidence of \eqref{PTfinal} with \eqref{schr-sw} requires
\begin{equation}
\frac{\tau_{t}}{y_{x}^{2}}=\frac{m_{0}}{m(t)} \, , \quad B(x,t)=-2\hbar \Omega(t)x \, .
\label{cond1}
\end{equation}
To simplify calculations we introduce real-valued functions $\mu(t)$ and $\sigma(t)$ such that $\tau_{t}=\sigma^{-2}(t)$ and $m(t)=m_{0}\mu^{2}(t)$. From the first condition in~\eqref{cond1} we get
\[
\tau (t)=\int^{t}\frac{dt'}{\sigma^{2}(t')}, \quad y(x,t)=\frac{\mu(t)x+\gamma(t)}{\sigma(t)}, 
\]
where $\gamma(t):\mathbb{R}\rightarrow\mathbb{R}$ emerges as a constant of integration with respect to $x$. The above result is quoted as Eq.~(\ref{TauY}) in the main text. On the other hand, the second condition in~\eqref{cond1} permits to determine the remaining transformation function $A(x,t)$, which we conveniently rewrite as 
\begin{equation}
A(x,t)=A_{0}(x,t)A_{1}(x,t) \, , 
\label{A1}
\end{equation}
with 
\[
A_{0}(x,t):=\exp\left[ i\tfrac{m_{0}}{\hbar}\tfrac{\mu}{\sigma}\left( \tfrac{1}{2} \mathcal{W}_{\mu} x^{2}+\mathcal{W}_{\gamma}x+\xi_{1}\right)\right] \, , 
\]
and
\[
A_{1}(x,t):=\exp\left[ -\tfrac{m_{0}}{\hbar}\mu^{2}\left( \Omega x^{2}+v x + \xi_{2} \right) \right] \, .
\]
The above expressions are quoted respectively as (\ref{miA2}) and (\ref{A2}). The real-valued functions $\xi_{1}(t)$ and $\xi_{2}(t)$ are integration constants with respect to $x$, and
\begin{equation}
\mathcal{W}_{\mu}\equiv\mathcal{W}_{\mu}(t)=\sigma\dot{\mu}-\dot{\sigma}\mu, \quad \mathcal{W}_{\gamma}\equiv\mathcal{W}_{\gamma}(t)=\sigma\dot{\gamma}-\dot{\sigma}\gamma \, , 
\label{miW}
\end{equation}
where $\dot f = \frac{df}{dt}$. In this form, $A(x,t)$ is factorized as the product of the gauge transformation $A_{0}(x,t)$, working as a unitary transformation, and the non-gauge term $A_{1}(x,t)$, accounting for the non-Hermiticity of $H_{\operatorname{sw}}(t)$.

The explicit form of the time-dependent potential $V(x,t)$ introduced in~\eqref{BV} is provided in Eq.~(\ref{V2}) of the main text, together with 
\begin{equation}
V_{0}(t)= m_{0} \left[ 
\tfrac{d}{dt}\left(\tfrac{\mu}{\sigma}\xi_{1}\right)-\tfrac{\mathcal{W}_{\gamma}^{2}}{2\sigma^{2}}-\tfrac{\mu^{2}v^{2}}{2}+\tfrac{w_{0}^{2}\gamma^{2}}{2\sigma^{4}} + \tfrac{\hbar}{m_{0}}\Omega 
\right] 
+ i m_{0} \left[ 
\tfrac{d}{dt}(\mu^{2}\xi_{2})-t\tfrac{\hbar}{2m_{0}} \tfrac{\mathcal{W}_{\mu}}{\sigma\mu} \right] \, .
\label{miV2}
\end{equation}

Comparing Eq.~\eqref{V2} with the potential associated to the operator~\eqref{Hsw2} gives a system of equations for $\sigma$, $\gamma$, $\xi_{1}$ and $\xi_{2}$. After making $V_0 = \hbar \Omega^{-1}$ one gets
\begin{equation}
\ddot{\sigma}+ \left[ 
w^{2}+4\Omega^{2}-\tfrac{\ddot{\mu}}{\mu}+\tfrac{2i}{\mu^{2}}\tfrac{d}{dt}(\mu^{2}\Omega) 
\right] \sigma=\tfrac{w_{0}^{2}}{\sigma^{3}} \, , 
\label{eqs1}
\end{equation}

\begin{equation}
\ddot{\gamma} + \left[
w^{2}+4\Omega^{2}-\tfrac{\ddot{\mu}}{\mu}+\tfrac{2i}{\mu^{2}}\tfrac{d}{dt}(\mu^{2}\Omega) 
\right] \gamma = 2v\mu\Omega-\tfrac{i}{\mu}\tfrac{d}{dt}(\mu^{2}v) \, , 
\label{misAeqs1}
\end{equation}
and
\begin{equation}
\frac{d}{dt}\left(\frac{\mu}{\sigma}\xi_{1}+i\mu^{2}\xi_{2}-\frac{\gamma}{2\sigma} \mathcal{W}_{\gamma} -i\frac{\hbar}{2m_{0}}\ln\frac{\mu}{\sigma} \right) 
- i\frac{\gamma}{2\mu}\frac{d}{dt}\mu^{2}v - \frac{\mu^{2}v^{2}}{2}+v\gamma\mu\Omega=0.
\label{miseqs1}
\end{equation}

\end{document}